%% file: MWAP.tex
\documentclass[conference]{IEEEtran}
\usepackage{cite}
\usepackage{amsmath,amssymb,amsfonts}
\usepackage{algorithmic}
\usepackage{graphicx}
\usepackage{textcomp}
\usepackage{xcolor}
\usepackage{framed}
\usepackage{textpos}
\def\BibTeX{{\rm B\kern-.05em{\sc i\kern-.025em b}\kern-.08em
    T\kern-.1667em\lower.7ex\hbox{E}\kern-.125emX}}

\usepackage{listings}
\usepackage{booktabs}
\usepackage{url}

\begin{document}
\title{Messaging with Purpose Limitation -- Privacy-Compliant Publish-Subscribe Systems}

\author{\IEEEauthorblockN{Karl Wolf}
\IEEEauthorblockA{\textit{Information Systems Engineering} \\
\textit{TU Berlin}\\
Berlin, Germany \\
kw@ise.tu-berlin.de}
\and
\IEEEauthorblockN{Frank Pallas}
\IEEEauthorblockA{\textit{Information Systems Engineering} \\
\textit{TU Berlin}\\
Berlin, Germany \\
fp@ise.tu-berlin.de} \and \IEEEauthorblockN{Stefan Tai}
\IEEEauthorblockA{\textit{Information Systems Engineering} \\
\textit{TU Berlin}\\
Berlin, Germany \\
st@ise.tu-berlin.de}}

\maketitle

\begin{textblock*}{\textwidth}(0cm,-8cm) %
\begin{center}
\begin{framed}
    \textit{Preprint version (2021-10-29), accepted as full paper at the 2021 IEEE Conference on Enterprise Computing (EDOC). \copyright2021 IEEE, to be published on IEEE Xplore}
\end{framed}
        
\end{center}
\end{textblock*}

\lstset{basicstyle=\ttfamily\footnotesize,breaklines=true}

\begin{abstract}
Purpose limitation is an important privacy principle to ensure that personal data may only be used for the declared purposes it was originally collected for. Ensuring compliance with respective privacy regulations like the GDPR, which codify purpose limitation as an obligation, consequently, is a major challenge in real-world enterprise systems.
Technical solutions under the umbrella of purpose-based access control (PBAC), however, focus mostly on data being held at-rest in databases, while PBAC for communication and publish-subscribe messaging in particular has received only little attention. In this paper, we argue for PBAC to be also applied to data-in-transit and introduce and study a concrete proof-of-concept implementation, which extends a popular MQTT message broker with purpose limitation. On this basis, purpose limitation as a core privacy principle can be addressed in enterprise IoT and message-driven integration architectures that do not focus on databases but event-driven communication and integration instead.

\end{abstract}


\input{MWAP_01_Introduction}
\input{MWAP_02_Background}
\input{MWAP_03_Requirements}
\input{MWAP_04_Considerations}

\input{MWAP_05_Implementation}

\input{MWAP_06_Evaluation}

\bibliographystyle{IEEEtran}
\bibliography{ma-zotero}

\end{document}

%% file: MWAP_01_Introduction.tex
\section{Introduction}

Privacy is of key importance to any enterprise. Regulations like the GDPR \cite{theeuropeanparliamentRegulationEU20162016} in the EU specifically prescribe how personal information is (not) to be used by organizations. Among the core principles of regulations like the GDPR is the principle of purpose limitation. This privacy principle %
requires purposes for the processing of data to be specified and that compliance to declared purposes must be ensured. 

Related work on purpose limitation, however, has mostly focused at data-at-rest, that is, data persisted in a database and accessed (for a particular purpose) later on. Respective technical mechanisms are typically subsumed under the term “purpose-based access control (PBAC)”. Concrete implementations proposed include low-level database extensions often referred to as “Hippocratic Databases” \cite{agrawalHippocraticDatabases2002,laura-silvaRealizingPrivacyPreservingFeatures2007,padmaHippocraticPostgreSQL2009} as well as higher-level approaches integrating PBAC into established programming abstractions for database access such as object-relational-mappers (ORMs) \cite{pallasApplicationLayerPurposeBasedAccess2020}.

Surprisingly, little attention has been paid on purpose limitation and PBAC for data-in-transit, that is, data traveling through the network by means of communication middleware. With event-driven architectures and stream-based processing, data may not necessarily be persisted anymore, but runs through processing pipelines in the form of messages. Especially in enterprise IoT settings, with large-scale integration of multiple data sources and use cases involving real-time monitoring, such message-oriented processing patterns are increasingly prevalent. 

To close this gap, we herein propose PBAC for data being in-transit in message-oriented architectures. In this regard, we aim to answer the following research questions: 

\begin{itemize}
    \item How can the privacy principle of purpose limitation be technically implemented in real-world message-oriented architectures? 
    \item What design options need to be considered in doing so? 
    \item What is the performance impact to be expected from the different design options?
\end{itemize}

To address these questions and the aforementioned gap, we introduce and study a concrete implementation, HivePBAC, an extension to one of the most widely used MQTT message brokers – HiveMQ. Our solution brings purpose-awareness to the publication and consumption of personal data in use cases following the publish-subscribe pattern. As such, our technical solution complements any administrative and organizational %
privacy measures common in practice.

In particular, our contributions are:

\begin{itemize}
    \item An in-depth analysis of requirements and prerequisites to be taken into account when introducing purpose-awareness to publish-subscribe systems,
    \item A model and design to introduce purpose-related functionalities -- particularly allowing the publication of data to be bound to sets of hierarchically structured allowed (AIP) and prohibited intended purposes (PIP) and subscriptions to be made for explicitly specified access purposes (APs) -- to existing MQTT brokers while still providing backward compatibility for clients without dedicated purpose-related functionality,
    \item A proof-of-concept implementation for a purpose-aware MQTT broker, enforcing said purpose restrictions against APs at different filtering points in time (on publish, on subscribe, etc.),
    \item An easy-to use Python client library allowing to integrate respective purpose-related capabilities into sending and receiving clients with low effort, and
    \item An initial performance evaluation of the proposed broker extension and its various modes of operation allowing to extrapolate realistic enterprise load patterns.
\end{itemize}

In the remainder of this paper, these contributions unfold as follows: In section \ref{background}, we provide the relevant background knowledge on purpose limitation, purpose-based access control, and publish-subscribe systems that guided our system design. On this basis, section \ref{requirements} derives a set of requirements to be fulfilled which then serve as the basis for our fundamental design considerations laid out in section \ref{design_considerations}. In section \ref{implementation}, we then present our proof-of-concept implementation and its different operation modes addressing different use-cases to be expected in practice. This implementation is subsequently subjected to an initial performance assessment (section \ref{evaluation}). Limitations and valuable subjects for future work are identified in section \ref{limitations}, section \ref{conclusion} concludes.

%% file: MWAP_02_Background.tex
\section{Background and Related Work}\label{background}

Our approach rests upon background knowledge and related work on purpose limitation in general, on respective technical implementations for purpose-based access control, and on publish-subscribe %
systems. All three pillars shall thus be briefly elaborated on.

\subsection{Purpose Limitation}

As already stated above, purpose limitation is one of the core principles underlying modern privacy regulations. The GDPR \cite{theeuropeanparliamentRegulationEU20162016} -- which increasingly serves as a blueprint for many comparable legislations around the world -- for instance prescribes that personal data must be ``collected for specified, explicit and legitimate purposes and not further processed in a manner that is incompatible with those purposes'' (Art. 5, 1b). In addition, purpose specification and limitation also apply to further concepts from the GDPR such as individually given consent which, in order to be effective, must also refer to ``one or more specific purposes'' (Art. 6, 1a) or the principle of data minimization, which states that the amount of processed personal data must be limited to ``what is necessary in relation to the purposes for which they are processed'' (Art. 5, 1c). The outstanding relevance of purpose limitation is also highlighted by its explicit mention with regard to ``privacy/data protection by design and by default'' in Art. 25 (2): Here, it is prescribed that controllers must ``implement appropriate technical and organisational measures for ensuring that [\ldots] only personal data which are necessary for each specific purpose of the processing are processed.''

The concept of purpose limitation was, however, not newly introduced with the GDPR in 2016. Instead, it was already anchored in the earliest data protection laws from the 1970s as well as in the GDPRs predecessor, the European Data Protection Directive of 1995 \cite{eu-dp-directive}. Similarly, non-binding frameworks and standards such as the Federal Trade Commission's Fair Informtion Practice Principles \cite{federal_trade_commission_privacy_1998}, the OECD's privacy framework \cite{oecd-guidelines}, or the ISO/IEC-standard 29100 \cite{iso-29100} explicitly or implicitly incorporate purpose limitation as one of the foundational principles of personal data privacy (see also \cite{pallasApplicationLayerPurposeBasedAccess2020}). Generally speaking, the privacy principle of purpose limitation is thus anchored in many regulations and non-regulatory frameworks that enterprises must comply with or adhere to -- be it to achieve legal compliance or to qualify for entering business relations that require conformance with respective standards -- when processing personal data.

\subsection{Purpose-Based Access Control (PBAC)}

So far, the principle of purpose limitation has mostly been implemented through rather administrative procedures ("privacy by policy", according to Spiekermann and Cranor~\cite{spiekermann2008engineering}), sometimes supported by combining several technical mechanisms not explicitly designed with purpose limitation in mind \cite{colesky2016critical}. Given the relevance of purpose limitation and the growing regulatory emphasis on implementing privacy principles ``by design'', however, several proposals for more explicit and thoroughly tailored technical materializations have been made in the past. In particular, this has been done under the terms of ``hippocratic databases'' and ``purpose-based access control''. Byun et al. \cite{byunPurposeBasedAccess2005,byunPurposeBasedAccess2008} were among the first presenting a formalized model %
and a prototypical implementation for the Oracle database. Subsequently, the same abstract concept was also applied to other relational databases such as IBM DB2 \cite{lefevreLimitingDisclosureHippocratic2004}, PostgreSQL \cite{laura-silvaRealizingPrivacyPreservingFeatures2007}, or Microsoft SQL \cite{AzemovicDataPrivacySQL2012} as well as to the document-based NoSQL database MongoDB \cite{colomboEnhancingMongoDBPurposeBased2017}. 

In medical or fitness-related use-cases, for instance, such schemes can be used to allow access to a given piece of health data ``for the purpose of individual treatment'' while access for a purpose like ``drug research'' may be blocked. In a car-sharing use-case, in turn, usage data of a particular driver might be made accessible for the purpose of billing but not for global route optimization or individualized advertisements, while other drivers' data might very well be used for advertising purposes. Insofar, purpose-based schemes provide a valuable complement to other access control models in implementing the principle of purpose limitation ``by design''.

All above-mentioned implementations proposed so far were, however, closely coupled to one particular database system, limiting the free choice of the database and the possibility to switch between different ones in practice. To attenuate respective lock-ins and increase flexibility, proposals for database-independent implementations of PBAC \cite{pallasApplicationLayerPurposeBasedAccess2020} have also been made. 

Still, these do also only cover use-cases where the data to be considered resides ``at rest'' in a database and is accessed from there, leaving other, increasingly relevant system classes uncovered by practically applicable implementations for purpose-based access control. In particular, this is so far the case for message-oriented middleware systems forming the backbone of modern, event-driven or stream-based enterprise architectures. In such architectures, the data to be considered are often not persisted at all but rather processed and accessed on the fly while being ``in transit'' through a larger processing pipeline. In this context, the class of publish-subscribe messaging systems addressed herein is of particular importance. These shall therefore briefly be introduced below.

\subsection{Publish-Subscribe and Access Control}

Publish-subscribe is a well-known, long-established communication paradigm that aims to decouple senders and receivers of data in distributed systems. Rather than sending data to specific receivers, publishers send messages to a broker, which passes them to interested subscribers based on subscription patterns \cite{eugsterManyFacesPublish2003}. Basically, publish-subscribe systems can follow topic-based or content-based patterns, whereas topic-based means that messages are sent to one or multiple topics organized in a hierarchical tree structure. In above-mentioned car-sharing use-cases, for instance, a car can repeatedly publish measured speed values to a topic \texttt{country1/area3/vehicle2342/speed}. Subscribers can then subscribe to one or multiple of these topics, whereas some systems allow the use of wildcards for receiving messages from any topic matching them (e.g., subscribing to \texttt{country1/area3/+/speed} could allow to receive all speed values measured in area 3). Widely used implementations of this topic-based approach include a multitude of brokers using the lightweight MQTT (Message Queuing Telemetry Transport) protocol \cite{andrewbanksMQTTVersion2019}, more feature-heavy approaches such as Apache Kafka \cite{apachesoftwarefoundationApacheKafka}, commercial products like IBM MQ \cite{ibm-mq}, or cloud-based solutions such as Amazon AWS SNS \cite{amazon-sns}. 

In content-based publish-subscribe systems, in turn, each message has a number of attributes and subscriptions are specified by a combination of conditions on these attributes. Such content-based subscriptions can require exact matches of a text attribute, an upper/lower bound of a number attribute, or even complex, query-like combinations evaluating multiple parameters \cite{eugsterManyFacesPublish2003}. For instance, this would allow receivers to subscribe to all messages from sensors of a certain assembly line which contain measured temperature values above a given value. Examples for such content-based systems include Hermes %
\cite{pietzuchHermesDistributedEventbased2002} and Siena \cite{carzanigaDesignEvaluationWideArea}. %

Their significantly more powerful, query-like capabilities notwithstanding, content-based publish-subscribe systems have so far not grown out of the primarily academic sphere and broadly established in enterprise practice. In contrast, topic-based ones are, due to their simplicity, their low computational footprint allowing for high-throughput applications, their fit to many real-world use-cases and, last but not least, the existence of standardized protocols like MQTT widely used in large-scale real-word applications. This is particularly true for IoT scenarios, many of which concern personal data, ranging from health telemetry \cite{ibmTelemetryUseCase2021} to personal car sharing \cite{hivemqgmbhCaseStudyBMW}. For our endeavor of bringing purpose-based access control to publish-subscribe systems, we therefore concentrated on topic-based ones using the MQTT protocol.

In matters of access control, the MQTT protocol provides capabilities for client authentication based on a combination of username and password, with advanced techniques based on a challenge-response scheme being added with version 5 \cite{andrewbanksMQTTVersion2019}. On this basis, different brokers implement different forms of access control for publishers as well as subscribers, ranging from simple access control lists (ACLs) \cite{light2017mosquitto} to more complex role-based access control (RBAC) models \cite{hivemqgmbhAuthorizationHiveMQDocumentation} or the integration of external APIs \cite{perroneDayMiraiSurvey2017,MosquittoConfMan2020,UserGuideEMQ}. %
However, all these schemes follow rather established access control models that do not account for the concept of purpose and the necessity to incorporate it into real-world access control mechanisms in the form of purpose-based access control. In the following, we therefore aim to establish a model and design as well as a proof-of-concept-implementation for coherently integrating PBAC into MQTT-based publish-subscribe systems, starting with the identification of requirements that need to be fulfilled.

%% file: MWAP_03_Requirements.tex
\section{Requirements}\label{requirements}

To provide a practically viable approach for technically materializing the concept of purpose limitation in real-world publish-subscribe systems, the integration model to be proposed must fulfill a couple of functional requirements. In addition to these, the design as well as the concrete implementation to be provided herein must also pay regard to a set of non-functional requirements. Both shall thus be laid out below.

\subsection{Functional requirements}

As a first, minimal set, we can particularly identify the following functional requirements that have to be met by any purpose-aware publish-subscribe system to be applicable and valuable in real-world scenarios: 

\paragraph{Capability to Handle Purposes on Publishing and Subscribing Side (Req. 1)} First and foremost, any mechanism for implementing PBAC in publish-subscribe systems must be capable of handling purposes on both sides of message-based communication. On the publishing side, it must be possible to specify the \emph{intended purposes} for which messages may be used and on the subscribing side, the particular, single \emph{access purpose} served by a subscription must be declarable. Only when these match, messages may be delivered.

\paragraph{Hierarchically Organized Purposes (Req. 2)} To allow for sufficiently specific purposes for fulfilling legal requirements, to ease the specification of more complex sets of intended purposes, and to heighten comprehensibility, a hierarchical organization of purposes has been widely established in the PBAC domain \cite{byunPurposeBasedAccess2008,pallasApplicationLayerPurposeBasedAccess2020}. The mechanism to be proposed should thus support hierarchically nested purposes with different levels of specificity.

\paragraph{Client-Driven Specification of Purposes (Req. 3)} To allow publishing and subscribing clients to use the mechanism to be proposed as flexibly as possible, intended purposes for published messages as well as the access purpose of a given subscription must be specifiable from the client side. Alternative approaches of centralized purpose management (e.g., with a central instance creating dedicated topics for certain purposes) are not considered viable in the context of highly distributed, loosely coupled, and message-oriented architectures.

\paragraph{Adaptability to Changed Intentions (Req. 4)} As intended purposes on the publishing side may change -- e.g. due to a withdrawal of consent (as a ``data subject shall have the right to withdraw his or her consent at any time'' \cite[Art. 7]{theeuropeanparliamentRegulationEU20162016}) -- such changes must be possible. %

\subsection{Non-functional requirements}

Beyond functional ones, several non-functional requirements must also be met in order to foster practical adoption of our design and implementation. Building upon previous works on practically applicable privacy engineering \cite{pallasApplicationLayerPurposeBasedAccess2020,gruenewald2021,ulbrichtYaPPLLightweightPrivacy2018}, we can here at least identify the following ones: 

\paragraph{Backward Compatibility (Req. 5)} The integration model and design shall be compatible to as many applications and clients as possible, with special regards to IoT devices, embedded systems, or legacy clients that may offer little or no way to adapt the way they publish or subscribe to messages. Ideally, both publishing and subscribing should thus be possible in a purpose-aware manner without changing anything on the client side. 

\paragraph{Extension of Existing Solution (Req. 6)} For the proof-of-concept implementation, an existing, sufficiently mature, and established solution should be built upon instead of creating a completely new broker from scratch. Doing so allows to concentrate on purpose-related functionalities alone while profiting from existing capabilities for message brokering in general etc. In addition, building upon a pre-existing solution eases practical applicability.

\paragraph{Reasonable Overhead (Req. 7)} Regulatory obligations to employ technical privacy %
measures are typically tied to the ``cost of implementation'' \cite[Art. 25]{theeuropeandataprotectionsupervisorHistoryGeneralData2016}, which primarily materialize in the form of (mostly performance) overheads. Similarly, such overheads will also be decisive for the application of privacy-related technologies such as PBAC beyond regulatory obligations. While some decrease in performance is to be expected, we therefore aim to keep it as low as possible to ensure practicability of the proposed mechanism and, consequently, its adoption. 

\paragraph{Developer Friendliness (Req. 8)} For the same reasons, the changes necessary to introduce purpose awareness into new or existing applications should be as little and intuitive as possible. This ensures both ease of integration and intuitive use of purposes in application code. %
In particular, a client-side library shall be provided to communicate purposes to the broker.

%% file: MWAP_04_Considerations.tex
\section{Design Considerations}\label{design_considerations}

Before going into the design and the prototypical implementation of a %
technical mechanism fulfilling the requirements identified above, several aspects %
require closer examination. Even though the details of different design choices and the implications thereof are better explicated in the light of a concrete implementation (see section \ref{implementation}), some preliminary, high level considerations shall be laid out first. %

\subsection{Expression of Purposes}
\label{cc:expression}

When expressing purposes, i.e. converting them from abstract real-life purposes to a formalized, automatically processable representation, %
the following aspects have to be considered:

\begin{itemize}
    \item \emph{Hierarchy}: While early technical approaches simply chose one or more purposes from a flat list \cite{p3p}, hierarchical organization (e.g., in the form of a tree \cite{byunPurposeBasedAccess2005} or a directed acyclic graph \cite{ulbrichtYaPPLLightweightPrivacy2018,pallasApplicationLayerPurposeBasedAccess2020}) of purposes allows to generalize or specialize as needed and enables a more concise notation, where allowing a purpose allows all respective sub-purposes without having to explicitly list each of them (see Req. 2). 
    \item \emph{AIP and PIP}: In addition to listing allowed intended purposes (AIP), specifying prohibited intended purposes (PIP) is a common approach in PBAC. Given a purpose tuple of (AIP, PIP), an access purpose (AP) is considered allowed if and only if AP itself or any parent purpose is in the AIP set AND neither AP nor any parent is in the PIP set. This makes notations of a large number of allowed purposes with a few prohibited child purposes significantly shorter, more readable and comprehensible \cite{byunPurposeBasedAccess2005} (again, see Req. 2).
    \item \emph{Mapping}: Rather than directly listing purposes as strings, purposes or even sets of purposes may be mapped e.g. to numerical IDs such as in \cite{pallasApplicationLayerPurposeBasedAccess2020}. This allows for shorter transmissions as well as a more coherent use of purposes across different involved components, but requires a prior, and potentially ongoing, synchronization of available purposes and their identifiers between the broker and every single client.
    \item \emph{Existing Languages}: Established concepts for formalizing privacy-related concepts such as consent or transparency \cite{ulbrichtYaPPLLightweightPrivacy2018,gerlModellingPrivacyLanguage,gruenewald2021} already include purposes and could be adopted. However, the overhead for doing so is significant. This might still be an option when other features of these languages, such as details on consent, retention times, etc. are needed anyway or when respective languages are already used in other parts of the overall application.
\end{itemize}

In the light of these considerations, we propose to employ hierarchically organized purposes represented as tuples of allowed and forbidden purposes (AIP, PIP). To keep flexibility high and complexity low, we abstain from incorporating the concepts of mappings and rather complex pre-existing languages, at least for the moment.

\subsection{Scope of Purposes}
\label{cc:scope}

A key decision in database-focused PBAC solutions is the scope, or granularity, of purposes: Shall AIPs/PIPs be specified per table, per column, per record, or per attribute? The same question also applies to publish-subscribe systems, for the scope of both AIPs/PIPs and APs (see Req. 1). \emph{AIPs/PIPs} could be specified per message, allowing for maximum flexibility (Req. 3) but introducing significant overhead (and, thus, contradicting Req. 7). Alternatively, AIPs/PIPs might be specified per session, thus attaching AIPs/PIPs to all messages published by a particular client. This would, however, limit flexibility as well as applicability in use-cases where one client publishes different message streams for different purposes. Finally, AIPs/PIPs could be specified per topic (or multiple topics through the use of wildcards) which is idiomatic for MQTT topics since topics group similar types of content that might consequently have the same set of AIPs/PIPs, thus only slightly reducing effective flexibility (Req. 3). %
We thus propose to specify AIPs/PIPs on a per-topic basis. 

Analogously, \emph{APs} might be attached to subscriptions (which in turn may entail multiple topics), sessions (assuming that any client only processes data for a certain purpose) or per topic. Here, a subscription-focused approach appears most reasonable.

For tying AIPs/PIPs to topics, we propose the client-driven (Req. 3) concept of \emph{reservations}: by reserving a topic, a client attaches a set of AIPs/PIPs to a topic filter, thus applying these AIPs/PIPs to all messages published to the concerned topic(s). This way, AIPs/PIPs have to be specified only once, and it is possible to perform reservations and subsequent message publications from different clients. This, in turn, can be used to implement purpose-awareness even for publishing clients not aware of the enhanced functionality and syntax, providing backward compatibility (Req. 5). As topic filters might overlap, \emph{effective intended purposes (EIP)} need to be computed from the combination of multiple reservations, as detailed in section \ref{cc:eip} below.

Similar to reservations, we propose \emph{presubscriptions} as backward-compatible (Req. 5) means to specify an AP before an actual subscription for a given client and topic; this allows non-purpose-aware clients to be subscribed purpose-fully using only the default MQTT protocol.

\subsection{Transmission of Purposes}
\label{cc:transmission}

Another design decision regards how purpose information is to be transmitted between the involved components communicating over MQTT, i.e., how purpose-related communication is to be integrated into the MQTT protocol while preserving backward compatibility (Req. 5). Basically, MQTT messages exist for three concepts: connection (and disconnection), subscription (and unsubscription) and publishing a message. Additional messages are used to keep the connection alive and ensure the selected level of reliability.

A typical MQTT message consists of the following elements: A set of fixed headers, variable headers, and -- for most message types -- a topic and payload. All of these message elements bear the potential for carrying purpose information. If necessary, dedicated messages to control purpose-related matters could also be used. %

Many existing approaches to enhancing MQTT opted for encoding data in \emph{fixed headers}, using potentially reserved fields for custom content \cite{bryceMQTTGPublishSubscribe2018,singhSecureMQTTInternet2015}. This has the benefit of not increasing the packet size, but is generally incompatible with brokers and clients unaware of the modified semantics and might conflict with other enhancements. MQTT version 5 allows the inclusion of \emph{user properties} in the variable header of most packet types \cite{andrewbanksMQTTVersion2019}. While providing an elegant and idiomatic way of including purpose information, this approach is, however, not supported by older MQTT versions, hindering our goal of backward compatibility (Req. 5). 

Storing data in the \emph{payload} -- e.g. by converting it to a JSON object containing the original payload as well as purpose-related metadata --  has the advantage of not changing the inner workings of the underlying MQTT protocol. After processing it, the broker could revert the payload into its original contents if necessary. However, only publish messages contain a payload, requiring an additional solution for embedding access purposes into subscriptions.

Finally, using \emph{topics} as means of transmitting purpose metadata would provide a consistent method for all message types and compatibility with all clients. This approach has been successfully used in MQTT+ \cite{giambonaMQTTEnhancedSyntax2018} with an extended subscription syntax such as \texttt{\$EQ;value/topic} for messages in \emph{topic} whose value equals \emph{value}. Given the advantages in matters of backward compatibility and coherent integration with the MQTT protocol, we thus propose to employ the topic structure for transmitting purposes.

\subsection{Determination of Compatibility}
\label{cc:eip}

The specification of AIPs, PIPs, and APs has been discussed in the previous sections. On this basis, the final compatibility decision whether a message is to be delivered or not (Req. 1) needs to be made. %
This, in turn, requires determining the effective intended purposes (EIPs) for a given topic. Three aspects need to be taken into account in this regard: 

First, \emph{combined} intended purposes, i.e. the merged purposes of all reservations whose topic filter matches the given topic. Multiple options of combining AIP/PIPs are conceivable, the most straightforward one being to take the union of all reservations' AIPs and PIPs respectively: A purpose is allowed if present in \emph{any} reservation's AIPs and forbidden if present in \emph{any} reservation's PIPs. 

Second, in the case of determining the compatibility of a wildcard subscription, \emph{affected} intended purposes need to be considered: subtopics might have different intended purposes. To prevent incompatible messages to be delivered through the subscription, all affected topics' AIPs and PIPs must be combined restrictively, i.e. only topics allowed in \emph{all} subtopics may be contained in the EIPs.

The resulting EIPs are once again a (AIP, PIP) tuple; An access purpose is considered compatible if and only if the purpose or a parent purpose is contained in the AIP set, but not the PIP set. %

\subsection{Filtering Point of Time}
\label{cc:fmodes}

Different from database-oriented PBAC approaches, message-oriented systems call for closer considerations about \emph{when} to actually perform the purpose-related filtering, i.e. determining EIPs and AP and deciding on the compatibility of the two. Two major paradigms are conceivable:

\emph{Filtering on publish}, i.e. checking every outgoing message: Which subscriptions match this message and for what AP were they created? Is one of them compatible with the messages topic's EIPs? If not, the message must not be delivered to this client.

\emph{Filtering on subscribe}, i.e. only allowing those subscriptions in the first place whose AP match their affected topics' EIPs. This poses the problem of partially allowed subscriptions when a wildcard subscription entails both compatible and incompatible topics while a subscription can only be either allowed or forbidden. In addition, existing subscriptions have to be re-evaluated when AIPs/PIPs change through a new or updated reservation (see Req. 4). Restricting the subscription to the allowed subset of topics on the broker side (as done in  \cite{belokosztolszkiRolebasedAccessControl2003}) is not feasible in MQTT, as it  would require knowing all used topics in advance. %
Rephrasing a wildcard subscription to multiple, more concrete subscriptions on the application side could, however, mitigate this issue.

As both approaches have their up- and downsides in matters of flexibility, expectable performance overheads under different load profiles, etc., we deem both basically valuable to be supported. The implementation chapter goes into more detail of the filtering modes as well as proposing a hybrid mode (see section \ref{imp:fmodes}).

%% file: MWAP_05_Implementation.tex
\section{Implementation}\label{implementation}

Our prototypical implementation -- HivePBAC -- is an extension to the HiveMQ broker that allows purpose-based communication in MQTT. We chose HiveMQ %
as the base system for our implementation for various reasons: First, it is fully compliant to the MQTT specification and supports both version 3 and 5 as well as TLS and web sockets to allow for backward compatibility (Req. 5). Second, HiveMQ is broadly used in enterprise practice and updated regularly, ensuring practical relevance (see Req. 6). Finally, it is open-source and provides a dedicated extension SDK.

Through this extension SDK, HiveMQ provides multiple ways for integrating custom behaviour into the broker: %
Interceptors, allowing to react to and change virtually any incoming or outgoing message; Authorizers, which can be used to allow or forbid connections and subscriptions as well as to set per-client and per-topic permissions for subscriptions and message publications; Authenticators check client's credentials on connection; and services allow to access and modify core aspects of the system, such as the list of subscriptions.

HivePBAC is implemented as such an extension and can be easily included in any running HiveMQ v4 broker by being inserted in its extension directory, thus causing minimal integration effort (Req. 8). In its default configuration, HivePBAC does not interfere with existing topics and messages (which do not have AIPs/PIPs) to ease a step-wise integration of purpose limitation into existing systems in a backward-compatible form (see Req. 5). However, it also provides a configuration option to strictly enforce purpose-aware usage. We provide HivePBAC freely and open source.\footnote{see https://github.com/PubSubPBAC} 

\subsection{Handling of Purposes}
\label{imp:purposes}

HivePBAC handles purposes on a \emph{per-topic} scope through the concept of \emph{reservations}: By sending a dedicated reservation message, a client ties a set of AIP/PIPs to a topic or topic filter (see Req. 3). %
In a reservation message, a filter allows to specify AIP/PIPs for more than one topic, or to specify general rules for a set of topics and individually allow or forbid further purposes for more specific topics. By sending another reservation for the same topic filter, AIP/PIPs for that filter are overwritten. Sending an empty reservation (as opposed to an empty AIP set!) removes the reservation for that topic string.

APs are handled on a  \emph{per-subscription} basis. Each subscription is created for a single, specific access purpose. To accommodate backward-compatibility with purpose-unaware clients (Req. 5), we additionally introduce the concept of \emph{presubscriptions}: A presubscription is a dedicated message to the broker, specifying a client ID, topic, and AP. HivePBAC stores that information, and applies the AP to later subscriptions with the given topic by the given client.

Keywords to signal a purpose-enhanced message, AIP/PIPs and AP are consistently transmitted inside message topics in the following format:

\noindent\texttt{!RESERVE/message/topic/or/filter\{AIP|PIP\}} \\
\texttt{!AP/message/topic/or/filter\{AP\}}

All keywords start with an exclamation mark in order to efficiently filter out non-keyword messages. In the prototypical implementation, keywords were chosen for expressiveness. However, all keywords (and characters, such as the curly brackets used to encode purposes in topics) mentioned in this work are centrally changeable to arbitrary strings supported by MQTT topics, i.e. any utf-8 string. In a production environment, it is recommendable to use keywords that are a) short and b) do not collide with any application-side topic names.

Picking up above-mentioned car-sharing use-case, access to a vehicle's location data may, for example, be restricted to the current driver's preferences upon a car-hire with the following reservation message:

\path{!RESERVE/country1/area3/vehicle2342/location/#{operational,marketing|marketing/individualized}}

The particular vehicle's loaction data may thus be subscribed to for all operational purposes (which might include billing as well as route optimization in the purpose tree) as well as for marketing, except  individualized marketing. Following this (AIP, PIP) specification, a subscription like \newline
\path{!AP/country1/area3/_/location/city{marketing/individualized}}
\newline would be blocked: While marketing as the parent purpose is allowed in the AIP list, the sub-purpose ``individualized'' is explicitly prohibited in the PIP list.

\subsection{Implementation Structure}
\label{imp:structure}

Our implementation rests upon a couple of interdependent building blocks: The \emph{reservation store} manages AIPs/PIPs per topic or topic filter. It is implemented as an interface to allow for different storage solutions, three of which we prototypically implemented. These are presented in section \ref{imp:stores}.
Likewise, the \emph{subscription AP store} stores APs per subscription or presubscription and is also implemented as an interface. HiveMQ already has a highly optimized subscription store for the actual subscriptions that can be accessed and changed through a service from the extension, but does not allow adding further information (as AP in this case) to a subscription. This necessitates to take the overhead of a second store. The core broker's subscription store is still interacted with to pause and unpause subscriptions if necessary, see section \ref{imp:fos}.

The \emph{command interceptor} checks incoming messages for commands, such as reservation, presubscription or requested changes of settings.
The \emph{subscription interceptor}, in turn checks incoming subscriptions for an access purpose. If present, the AP is stored in the subscription AP store and then removed from the subscription message's topic. The so-stripped subscription is then forwarded to the core broker, allowing it to be processed normally without interfering with the original routing algorithm. Analogously, an \emph{unsubscription Interceptor} removes subscription APs from the store upon unsubscription.

When filtering on subscribe, the \emph{subscription authorizer} is called by the broker to decide whether or not a subscription is allowed. It checks the compatibility of the subscription's AP and possible reservations for its topics (as fetched from the respective stores). If compatible, the subscription is allowed and the core broker handles it further.
When filtering on publish, the outbound \emph{publish interceptor}, finally, checks every outgoing message for the existence of a matching AND compatible subscription. If incompatible, the delivery of the message is prevented.

\subsection{Filtering Modes}
\label{imp:fmodes}

A key implementation decision is at what point of time the filtering ultimately happens. As described in section \ref{cc:fmodes}, two major options exist: filtering on publish, i.e. allowing or forbidding outgoing messages, or filtering on subscribe and, thus, allowing or forbidding subscriptions. This section describes each mode in more details and introduces a \emph{hybrid} mode that aims to combine some of the advantages of both modes. %

\subsubsection{Filter on Subscribe (FoS)}
\label{imp:fos}

When filtering on subscribe, Hive\-PBAC evaluates \emph{subscriptions} at the time of subscription, and whenever a condition of compatibility changes, i.e. upon reservation changing AIPs/PIPs (see Req. 4). 

As briefly touched in section \ref{cc:eip}, two cases need to be considered: \emph{combined reservations}, i.e. multiple sets of wildcard reservations that concern the subscription's topic and will have to be merged in order to derive the EIPs for that particular topic. In case of a susbscription with a wildcard topic filter, \emph{affected} topics need to be checked as well: Once the subscription is allowed, the broker will deliver all messages matched by the subscription's topic filter. To ensure purpose limitation, the AP of the subscription thus has to be compatible with \emph{all} topics matching the subscription's topic filter. All affected topics are thus combined AND-wise and intersected with the combined purposes. Only if the AP is compatible with these EIPs, the subscription can be authorized.

A similar process happens when a reservation changes: All subscriptions that the changed reservation is \emph{combined} in (i.e. that are entailed by its topic filter), and all subscriptions that are \emph{affected} by the reservation, (i.e. which the reserved topic filter is a subtopic of) need to be re-authorized. If a formerly allowed subscription is now incompatible, this can be handled in two ways: As the MQTT specification provides no way for retroactively forbidding subscriptions, the only way to notify the client is disconnecting it, with the downside of disrupting all other subscriptions of that particular client as well. Upon reconnection, the client has to redo all its subscriptions and can now transparently see which ones are allowed or forbidden. The other way, which is used in HivePBAC, is \emph{pausing} the subscription. By pausing it, it is removed from the core broker's subscription directory, preventing the broker from considering it for delivering messages. The subscription AP is kept and allows it to be \emph{unpaused} when another change of reservation renders the subscription allowed again. The subscription is then recreated in the core broker with the information from the subscription AP store. In this case, the client is not aware of not being authorized for the subscription anymore. %

The upside of filtering on subscribe is the low expected overhead: Once a client is subscribed, no more filtering decisions need to be made by the extension regardless of the number of messages sent through the subscription. The disadvantage is a higher subscription time and, more significantly, the missing ability to subscribe to a large number of topics that are only partially allowed: When only one single subtopic is incompatible, the entire subscription is denied or paused, rather than still sending the remaining, allowed subset of messages.

\subsubsection{Filter on Publish (FoP)}

When filtering on publish, all subscriptions are initially accepted and subscription APs are respectively saved to the AP store. The filtering, however, occurs for each outgoing message. At this point, the core broker has already found matching subscriptions and sends a message to each subscribed client (once per client, not per subscription). Unfortunately, the extension has no way of knowing based on which subscription(s) the message is being delivered, so it has to separately identify all matching subscriptions again and check each of them for purpose compatibility. In contrast to FoS, only combined reservations are relevant here, as a message always has a specified, single topic rather than a filter. If the combined EIP of that topic are compatible with \emph{any} of the matching subscriptions' AP, the message is published to the client. When a message is matched by multiple subscriptions, the client does not know on the basis of which subscription -- and, consequently, which AP -- a message is received. %

The natural downside of filtering on publish is that for many published messages, many checks have to be performed, resulting in a significant overhead to be expected. The upside is that subscriptions can be partially allowed, enabling fine-grained access control: A subscription can entail many subtopics and have the allowed subset delivered. Additionally, the expected overhead resulting from rapidly changing AIPs/PIPs or subscriptions is much lower, as no checks have to be performed at the time of reservation or subscription. %

\subsubsection{Hybrid Filtering (Hbr)}

The hybrid mode aims to combine the advantages of both approaches: Subscriptions that are \emph{fully} incompatible, i.e. their combined EIPs (from all reservations concerning the topic) do not include the subscription's AP, are forbidden, similarly to FoS. Partially (or entirely) compatible subscriptions are still allowed. Thus, affected subtopics of a wildcard subscription do not require checking on subscribe. Messages for allowed subscriptions are filtered on publish based on their actual topic, as in FoP. This enables partially allowed subscriptions as in FoP, but saves filtering overhead for fully forbidden subscriptions, as they are transparently forbidden at the time of subscribe. %

As no filtering type is objectively best, the prototypical implementation supports all three, allowing to choose via configuration. The (so far assumed) implications of these modes in matters of performance will be briefly evaluated %
in section \ref{evaluation}.

\subsection{Reservation and Subscription Stores}
\label{imp:stores}

As introduced in section \ref{imp:structure}, two major categories of \emph{purpose metadata} need to be held by the extension: reservations, representing allowed and prohibited intended purposes, and subscription APs, representing access purposes per subscriptions. The extension implements both of them as interfaces to allow for different implementations. Three approaches are implemented in the prototype with the potential for more optimized storage in the future.

The first, \emph{``flat''}, implementation na\"ively  builds both stores based on hash maps. For reservations, a set of AIPs is mapped to a topic filter as string. This makes storing and accessing individual reservations very fast, but does not allow to efficiently access logically connected reservations, i.e. finding all reservations for topics matched by a given filter. Analogously, subscription APs are held as SubscriptionAP object, containing topic filter, QoS level, and AP of a given subscription and are stored in nested hash maps per client ID per topic filter. The QoS level is not needed for access control per se, but allows to recreate a subscription in case it has been paused (see section \ref{imp:fos}). Finding combined or affected subscriptions or reservations requires checking all reservations or subscriptions for matching a wildcard topic; this is realized using basic string operations where possible and regular expressions were necessary. Due to the need to iterate many reservations / subscriptions and the use of computationally expensive operations, this mode of storage is not expected to scale well.

The \emph{tree-based} implementation aims to store purpose metadata in a more idiomatic way by mirroring the topic structure. Individual node objects represent topic levels, including wildcards, with each node storing a hash map of its subnodes, much like a prefix tree. Each node then stores AIPs and PIPs for the topic it represents (if set by a reservation) and its subscriptions. Finding combined and affected reservations or subscription is now a matter of recursively walking down the tree and is expected to scale much better with a large number of reservations and/or subscriptions. 

As purpose compatibility decisions for individual topics or topic filters do not change unless the underlying reservations change, it makes sense to cache these decisions. A basic \emph{cached store} wraps one of the above stores and caches the most relevant outputs for filtering in a hash map: For a given topic string, what are the effective intended purposes considering combined and affected reservations? For a given client and topic, does a \emph{matching} and \emph{compatible} subscription exist? In our first, simple implementation of this caching approach, the entire cache is invalidated when a reservation is created or changed. A future, more sophisticated approach might, however, only invalidate relevant parts of the cache at the cost of having to determine which parts are relevant.

To elucidate the actual impact of these different approaches, their performance is again briefly compared against each other in section \ref{evaluation_perf}.

\subsection{Client}
\label{cc:imp}

To ensure backward compatibility (Req. 5) and foster practical adoption (Req. 8), our approach is already designed to require as little changes on the client side as possible. Introducing purpose awareness is mostly a matter of adapting topics, following a straightforward pattern. Nevertheless, we also provide a python library natively allowing for the implementation of purpose-aware MQTT clients.\footnote{see https://github.com/PubSubPBAC} We build upon the widely established Eclipe Paho library \cite{eclipsefoundationEclipsePahoPython}. In order to ensure being able to profit from future developments, we again decided against merely changing this library but rather opted for a wrapping approach. Using this library, a developer intending to restrict access to all messages published under the \texttt{home/} path to the purposes \texttt{marketing (except analytics)} and \texttt{operational} and wanting to receive messages from all power sensors for the purpose \texttt{billing} (a sub-purpose of \texttt{operational}) can simply do so as follows:

\begin{lstlisting}[label=lst:pythonclient,language=Python,caption={A purpose-aware python client is set up using the Eclipse Paho client as a basis. After connecting to the MQTT server, the client makes a reservation, subscribes to a topic in a purpose-aware manner and sends a message.}]
import paho.mqtt.client as mqtt
from purpose_client import PurposeClient

mqtt_client = mqtt.Client(client_id="purpose_client", clean_session=True)
client = PurposeClient(mqtt_client)
client.connect("localhost", 1883, 60)

client.reserve("home/#", aip=["marketing", "operational"], pip=["marketing/analytics"])
client.subscribe_with_purpose("home/sensors/power/#", "operational/billing")

client.send("home/sensors/power/392/total", b"3142")
\end{lstlisting}

With purpose-limitation being implementable on the client with such small effort and a purpose-aware extension smoothly hooking into a widely established broker available and fulfilling all previously identified requirements, the additional implementation effort for bringing legally required purpose-awareness to real-world applications employing the publish-subscribe pattern appears bearable for many cases. However, practical adoption also requires bearable overheads in matters of performance (Req. 7). Our implementation shall therefore also be briefly examined in this regard.

%% file: MWAP_06_Evaluation.tex
\section{Preliminary Performance Evaluation}\label{evaluation_perf}
\label{evaluation}

The practical applicability of any privacy-related technology and, probably more importantly, data controllers' obligation to apply it under regulatory ``privacy/data protection by design'' provisions referring to the ``costs of implementation'' \cite[Art. 25]{theeuropeandataprotectionsupervisorHistoryGeneralData2016} strongly depends on the performance overhead it introduces. %
In addition, the impacts of different filtering modes under different settings were so far only roughly predicted, also calling for further validation. We therefore conducted respective preliminary experiments in matters of message throughput, message latency, executable subscriptions per timeframe, etc., depending on factors such as the number of active clients, the filtering mode, or the number of active -- and thus, to be resolved -- reservations.

Using a self-written tool based on an existing MQTT benchmark\footnote{\url{https://github.com/hui6075/mqtt-bm-latency}}, we were particularly interested in the overhead  on message throughput, latency, and the number of processible subscriptions our extension introduces compared to plain HiveMQ with the different filtering modes. We thus compared the unchanged HiveMQ broker as the default reference (\emph{Hive}) with a ``dummy'' filtering extension that only scans incoming messages for PBAC-related commands but performs no actual filtering (\emph{NoF}), filtering on subscribe without any caching (\emph{FoS}), filtering on publish without and with caching, using a tree-based AIP/PIP-store (\emph{FoP} and \emph{FoP\_Cache}), and, finally filtering on publish with a flat store for AIP/PIP-resolution (\emph{FoP\_Flat}). 

For our preliminary assessment, we deliberately chose comparably small numbers for active clients (up to 250), messages (up to 10.000), and subscriptions (3 per client) to allow for the closer examination of result changes. At the same time, the chosen numbers are high enough to show stable results or trends, suggesting it valid to extrapolate from them to higher numbers. All experiments were conducted in line with established best-practices for security- and privacy-related performance benchmarks \cite{Pallas2018Disillusion,pallasApplicationLayerPurposeBasedAccess2020,pallas-ea-evidence-based}

Basically, our results confirm our expectations from section~\ref{implementation}: From a stabilization at 100 publishing clients onwards, the ``dummy'' filtering extension (\emph{NoF}) consistently results in throughput reductions of 12-14\%, which highlights the non-negligible overhead of the extension systems and the command-filtering itself. %
Results for \emph{FoS}, where filtering doesn't have to be executed for every single message, were quite aligned to those for \emph{NoF} with differences being in the range of measurement uncertainties. The three filtering modes \emph{FoP}, \emph{FoP\_Cache}, and \emph{Hbr} resulted in quite comparable overheads with a measurable advantage resulting from caching. %
The results for filtering on publish with a flat (\emph{FoP\_Flat}) opposed to the tree-based one used in all other \emph{FoP} variants, in turn, vividly demonstrate the impact of respective optimizations laid out in section \ref{imp:stores} (see table \ref{tab:pubsuboverhead}).

\begin{table} %
  \centering
\begin{tabular}{lrr}
\toprule 
Mode &  Throughput (msg/s) &  Relative Overhead (\%) \\
\midrule
Hive      &     6254.49 &         --- \\
FoS       &     5467.28 &        12.59 \\
NoF       &     5404.29 &        13.59 \\
FoP\_Cache &     4998.86 &        20.08 \\
Hbr       &     4690.76 &        25.00 \\
FoP       &     4638.01 &        25.85 \\
FoP\_Flat  &     3519.83 &        43.72 \\
\bottomrule
\end{tabular}
  \caption{Exemplary throughput and relative overheads compared to vanilla HiveMQ at 250 concurrent clients}
  \label{tab:pubsuboverhead}
\end{table}

In matters of latency, we observed comparable relations (see table \ref{tab:pubsublatency}): \emph{FoS} and \emph{NoF} exhibit latencies increased by less than 10\%. The seemingly \emph{higher} latency of \emph{NoF} can here be attributed to the measurement variations typically seen in cloud environments. Of the different variants for filtering on publish, the cached one is again the most performant one, followed by the hybrid approach and the plain one. Again, the difference between \emph{FoP\_Cache} and \emph{Hbr} is more prominent than the one between \emph{Hbr} and \emph{FoP}. For \emph{FoP\_Flat}, the computational filtering overhead becomes more than apparent and materializes in more than 22-fold round-trip latencies.%

For the practical implementation of purpose-based access control in publish-subscribe scenarios, these results suggest especially \emph{FoS} to be a viable option, albeit coming with the drawbacks regarding partially allowed subscriptions mentioned in section \ref{imp:fmodes}. The fact that the mere scanning of incoming messages for commands alone leads to basically similar performance drops as \emph{FoS} suggests this mechanism to be the first target for further optimizations in the future. In the case of significant overhead reductions being possible here, this will also reduce the overhead to be expected for the -- functionally preferable, see again section \ref{imp:fmodes} -- three variants of filtering messages on publish with a tree-based store, presumably rendering them practically viable, too. Filtering on publish based on a flat store, however, seems rather impractical, especially for higher numbers of reservations.\footnote{This is also supported by benchmarking results for higher numbers of subscriptions not presented herein, where message throughputs for \emph{FoP\_Flat} dropped significantly, again confirming expectations from \ref{imp:fmodes}.}

\section{Limitations and Further Work}\label{limitations}

Even though our prototypical implementation already fulfills all functional and non-functional requirements identified in section \ref{requirements} and even if the performance overheads determined above do at least speak in favor of its fundamental applicability of technical mechanisms for enforcing the privacy principle of purpose limitation within message-oriented publish-subscribe architectures in practice, we can of course identify a couple of limitations and, respectively, promising subjects for future work.

One functional limitation of the state of implementation reached so far regards scalability. In use-cases with large amounts of messages, brokers are typically distributed across multiple instances. Such distribution would, however, require significantly more sophisticated approaches (e.g., for synchronizing across different instances) for implementing the purpose-related functionalities presented herein. Adding such mechanisms to our proof-of-concept implementation is thus an obvious target for future work.%

\begin{table} [t]
\centering
\begin{tabular}{lrrr}
\toprule
{Mode} &  Round-Trip Latency (msec) &  Relative Overhead (\%)\\
\midrule
Hive      &  50.64 &          --- \\
FoS       &  54.13 &          6.89 \\
NoF       &  55.20 &          9.01 \\
FoP\_Cache & 60.59 &         19.64 \\
Hbr       &  67.87 &         34.02 \\
FoP       &  68.95 &         36.16 \\
FoP\_Flat  & 1148.30 &       2167.57 \\
\bottomrule
\end{tabular}
  \caption{Exemplary round-trip latencies and relative overhead compared to vanilla HiveMQ at 250 concurrent clients}
  \label{tab:pubsublatency}
\end{table}

More in-depth performance benchmarks with message, client, and subscription numbers well beyond those used in our preliminary assessment should also be conducted to gain more insights about the system behavior to be expected in more complex scenarios, to avoid unforeseen surprises and to better identify existing bottlenecks introduced by our purpose-related extensions. Similarly, benchmarks for different usages of publish-subscribe systems (e.g., publication- vs. subscription-heavy, sparse vs. dense topic structures, etc.) would expectably provide highly valuable insights. Such advanced benchmarking efforts would, of course, go hand in hand with respective performance optimizations, which are particularly advisable for command-matching but also for other string-based operations currently used in our proof-of-concept implementation.

Besides extending our implementation for HiveMQ, we also foresee our design and implementation concepts to be adoptable to other topic-based message brokers widely used in practice with reasonable effort. Doing so would vividly demonstrate their generalizability and make purpose-aware messaging available for a broader variety of real-world applications and use-cases. 

Questions regarding the roles, needs, and support of all stakeholders involved in implementing the privacy principle of purpose limitation in practice (ranging from legal departments over systems architects to affected end-users) are also beyond the scope of this paper. Now that the technical feasibility of purpose-aware publish-subscribe systems has been demonstrated, such considerations are, however, a manifest target for follow-up research.

\section{Conclusion}\label{conclusion}

Motivated by legally given obligations for purpose limitation, the need to implement this privacy principle technically, and the lack of respective mechanisms for doing so in enterprise systems beyond and in addition to traditional databases, we herein presented a model and design for implementing purpose limitation in MQTT-based messaging brokers and architectures. %
Our prototypical proof-of-concept implementation, HivePBAC, provides advanced capabilities for purpose-aware publishing and subscribing based on hierarchical purpose structures, AIP/PIP combinations, and MQTT-specific topic patterns, implements different filtering modes (on publish, on subscribe, etc.) addressing different settings, and is implemented as a drop-in extension to a well-established MQTT broker to ensure low-effort adoptability and future-proofness. Preliminary performance benchmarks demonstrate the general suitability of our approach for being applied in practice while at the same time pointing towards several promising targets for future optimizations.

These so far unused optimization potentials and some further limitations notwithstanding, HivePBAC is -- to the best of our knowledge -- the first practically usable implementation of purpose-based access control in publish-subscribe systems. Our approach and implementation concepts should be adoptable for other topic-based message brokers with reasonable effort. %

For achieving legal compliance in matters of purpose limitation, enterprise systems, especially those following an event-driven and message-oriented model, can not rely on PBAC for data-at-rest only. This is particularly so in pace-gaining domains like large-scale IoT or real-time data integration. With solutions like HivePBAC, enterprises can technically ensure legal compliance in such settings where personal data are often processed ``on the fly'' instead of being persisted to and accessed from traditional databases.